 \documentclass[preprint2]{aastex}

\usepackage{color}

\newcommand{\bfm}[1]{\mbox{\boldmath $#1$}}
\newcommand{\mat}[1]{\mbox{\boldmath $\mathsf{#1}$}}
\newcommand{\pcm}{\,cm$^{-3}$}
\newcommand{\kms}{\,km\,s$^{-1}$}
\newcommand{\grad}{{\rm grad}}

\newcommand{\lesim}{\,\raisebox{-0.4ex}{$\stackrel{<}{\scriptstyle\sim}$}\,}

\shortauthors{Downes \& O'Sullivan}

\begin{document}


\title{Multifluid magnetohydrodynamic turbulent decay}



\author{T.P. Downes\altaffilmark{1,2,3} and S. O'Sullivan\altaffilmark{4}}
		
\email{turlough.downes@dcu.ie}

\altaffiltext{1}{School of Mathematical Sciences, Dublin City University,
	Glasnevin, Dublin 9, Ireland} 
\altaffiltext{2}{School of Cosmic Physics, Dublin Institute for Advanced
Studies, 31 Fitzwilliam Place, Dublin 2, Ireland}
\altaffiltext{3}{National Centre for Plasma Science and Technology,
	Dublin City University, Glasnevin, Dublin 9, Ireland}
\altaffiltext{4}{School of Mathematical Sciences, Dublin Institute of 
Technology, Kevin Street, Dublin 8, Ireland}

\begin{abstract}
It is generally believed that turbulence has a significant impact on the
dynamics and evolution of molecular clouds and the star formation which
occurs within them.  Non-ideal magnetohydrodynamic effects are known to
influence the nature of this turbulence.  We present the results of a suite of 
$512^3$ resolution simulations of the decay of initially super-Alfv\'enic and 
supersonic fully multifluid MHD turbulence.  

We find that ambipolar diffusion increases the rate of decay of the
turbulence while the Hall effect has virtually no impact.  The
decay of the kinetic energy can be fitted as a power-law in time and the
exponent is found to be $-1.34$ for fully multifluid MHD turbulence.
The power spectra of density, velocity and magnetic field are all
steepened significantly by the inclusion of non-ideal terms.  The
dominant reason for this steepening is ambipolar diffusion with the Hall
effect again playing a minimal role except at short length scales
where it creates extra structure in the magnetic field.

Interestingly we find that, at least at these resolutions, the majority of the 
physics of multifluid turbulence can be captured by simply introducing 
fixed (in time and space) resistive terms into the induction equation without 
the need for a full multifluid MHD treatment.

The velocity dispersion is also examined and, in common with previously
published results, it is found not to be power-law in nature.

\end{abstract}


\keywords{MHD — ISM: kinematics and dynamics — ISM: magnetic fields —
methods: numerical — turbulence}

\section{Introduction}
\label{sec:intro}

%
%

Turbulence is recognized as a possible source of support against
gravitational collapse for molecular clouds.  The precise role and
source of the observed motions interpreted as evidence of turbulence in
these clouds has been studied extensively by many researchers (see the reviews 
of \citealt{mac04,elm04}).  Clearly, if turbulence can support molecular
clouds then it can influence star formation in terms of rate, efficiency 
and initial mass function \citep{elm93,kle03}.

Many studies of turbulence in molecular clouds have focused on
ideal magnetohydrodynamics (MHD) as an approximation of the physics
governing this system \citep{maclow98, maclow99, ost01, ves03, gus06, 
glover07, lem08, lem09}.  The assumption of ideal MHD, while desirable for 
technical reasons, is perhaps risky in the context of turbulence.  The reason 
for this is that while ideal MHD is valid in molecular clouds on fairly large 
length scales, on shorter length scales non-ideal effects are thought to 
become significant \citep{wardle04, ois06}.  Given that turbulence in 3 
dimensions involves the transfer of energy from large scales to ever smaller 
scales, the assumption of ideal MHD will be invalid below some critical 
spatial scale and the correct nature of the energy cascade may not be observed 
at this range.

The most important of the non-ideal effects for molecular cloud dynamics is 
ambipolar diffusion. Some authors \citep{ois06, li08, kud08} have studied 
driven MHD turbulence in the presence of ambipolar diffusion.  All these 
authors find that ambipolar diffusion produces significant differences in the 
properties of the turbulence.

While most likely of lesser significance, it has been suggested that although 
the Hall resistivity is generally at least an order of magnitude lower than 
the ambipolar resistivity in molecular clouds \citep{wardle04}, its effect 
should not be ignored.  Although relatively weak, it is capable of inducing 
topological changes in the magnetic field which are quite distinct to any 
influence caused by ambipolar diffusion.  In support of this assertion, we 
note that researchers working on reconnection and the solar wind have studied 
the Hall effect in the context of turbulence and found that, although the 
overall decay rate appears not to be affected, the usual coincidence of the 
magnetic and velocity fields seen in MHD does not occur at small scales 
\citep{mat03, min06, ser07}.  Almost no work has been done on comparing the 
influences of this effect coupled with that of ambipolar diffusion on 
turbulence with the exception of \citet[hereafter Paper I]{dos09}. 

In Paper I a series of simulations of decaying supersonic non-ideal 
MHD turbulence incorporating both ambipolar diffusion and the Hall effect were 
performed.  These simulations, however, were constrained in that
the resistivities associated with each of ambipolar diffusion, the Hall
effect and the Pederson resistivity were kept fixed in both space
and time.  The authors found that, at length scales of 0.2\,pc, ambipolar 
diffusion has a significant impact on the decay of the turbulence.  The Hall 
effect was less significant in this respect but does have an influence on the 
magnetic field at short length scales.  Here we present simulations in which 
the resistivities are self-consistently calculated from the 
evolution of both the magnetic field and the densities of all of the 
component species of the fluid.  Using these dynamically evolving 
resistivities we study the decay of fully multifluid MHD turbulence.  This is 
the first such study presented in the literature, with the exception of the 
low resolution simulations presented by \cite{dos08}.

The aim of this work is to examine in detail the differences between the 
decay of ideal MHD turbulence and that of multifluid MHD turbulence with a 
full tensor resistivity incorporating the effects of ambipolar diffusion, the 
Hall effect and Ohmic resistivity.  We will use the results of Paper I in our 
discussion of these differences as it represents an intermediate stage between 
the calculations presented here and those of ideal MHD.  This work is new in 
two respects: notwithstanding Paper I, no previous work has focused on 
{\em decaying} (i.e.\ un-driven) {\em multifluid} MHD turbulence and, in 
addition, no previous work has addressed the issue of turbulence in the 
presence of both ambipolar diffusion and the Hall effect simultaneously.

%
%

In section \ref{sec:num-method} we outline the numerical techniques used
in this work, as well as the initial conditions and general set-up for the 
simulations while in section \ref{sec:analysis} we describe the methods
used to analyze the simulation data.  In section \ref{sec:results} we 
present and discuss the results of our simulations of turbulent decay.  
Finally, section \ref{sec:conclusions} contains a summary of our results.

\section{Numerical method}
\label{sec:num-method}
%
%

As in Paper I, we use the code HYDRA \citep{osd06, osd07} to integrate
the equations of weakly ionized multifluid MHD (see section 
\ref{subsec:eqns-algorithm}).  We assume that the molecular cloud material we 
are simulating can be treated as isothermal and that initially the density and 
magnetic field are uniform.  We use the capabilities of HYDRA to extend the 
physics incorporated in the simulations here beyond those presented in
Paper I so that the turbulence here is fully multifluid MHD. 

\subsection{Equations and algorithm}
\label{subsec:eqns-algorithm}
%
%

We briefly outline the equations and assumptions in our model here but
refer the reader to \cite{osd06, osd07} for a comprehensive description
of the underlying assumptions for the weakly ionized model of multifluid
MHD.

We assume that the cloud material can be treated as weakly ionized.  This is 
clearly valid for molecular clouds and allows us to ignore the inertia of the 
charged species \citep{cio02, falle03}.  For a system composed of $N$
fluids, one of which is neutral, the equations to be solved
are then
\begin{eqnarray}
\frac{\partial \rho_i}{\partial t} + \bfm{\nabla} \cdot \left(\rho_i
		\bfm{q}_i\right)  =  0 \mbox{, ($1 \leq i \leq N$), }
\label{mass} \\
\frac{\partial \rho_1 \bfm{q}_1}{\partial t}
	+ \nabla\cdot\left( \rho \bfm{q}_1 \bfm{q}_1 + a^2\rho\mat{I}\right) 
	= \bfm{J}\times\bfm{B} , \label{neutral_mom} \\
\frac{\partial \bfm{B}}{\partial t} + 
\nabla\cdot(\bfm{q}_1\bfm{B}-\bfm{B}\bfm{q}_1)  = 
-\nabla\times\bfm{E}' , \label{B_eqn} \\
\alpha_i \rho_i\left(\bfm{E} + \bfm{q}_i \times \bfm{B} \right) +
		\rho_i \rho_1 K_{i\,1}(\bfm{q}_1-\bfm{q}_i) = 0\label{charged_mom}\\
\nabla\cdot\bfm{B} = 0 \label{divB} , \\
\nabla\times\bfm{B} =  \bfm{J} , \label{eqn-J} \\
\sum_{i=2}^N \alpha_i \rho_i  =  0 , \label{charge_neutrality}
\end{eqnarray}
where $\rho_1$, $\bfm{q}_1$, $a$, $\bfm{B}$ and $\bfm{J}$ are the neutral
mass density, neutral velocity, sound speed, magnetic field and current
density respectively and $2\leq i \leq N$ unless otherwise noted.  
$K_{i\,1}$, $\alpha_i$ and $\rho_i$ ($i > 1)$ are
the collision coefficients between species $i$ and the neutrals, the charged 
fluid charge-to-mass ratios and mass densities respectively.  Equations
(\ref{mass}) to (\ref{charge_neutrality}) express conservation of mass
for each fluid, conservation of neutral momentum, the induction
equation, force-balance for the charged species, the inadmissibility of
magnetic monopoles, Faraday's law and charge neutrality respectively.

The electric field in the frame of the fluid, $\bfm{E}'$, is calculated from 
the generalized Ohm's law for weakly ionized fluids (e.g. \citealt{falle03, 
osd06}) and is given by
\begin{equation}
\bfm{E}' = \bfm{E}_{\rm O} +\bfm{E}_{\rm H} +\bfm{E}_{\rm A} ,
	\label{eqn_E1}
\end{equation}
where
\begin{eqnarray}
\bfm{E}_{\rm O} & = & (\bfm{J}\cdot\bfm{a}_{\rm O})\bfm{a}_{\rm O} ,
	\label{eqn_EO1} \\
\bfm{E}_{\rm H} & = & \bfm{J}\times\bfm{a}_{\rm H} , \label{eqn_EH1} \\
\bfm{E}_{\rm A} & = & -(\bfm{J}\times\bfm{a}_{\rm A})\times\bfm{a}_{\rm A}
	, \label{eqn_EA1} 
\end{eqnarray}
using the definitions $\bfm{a}_{\rm O}\equiv f_{\rm O} \bfm{B}$,
$\bfm{a}_{\rm H}\equiv f_{\rm H} \bfm{B}$, $\bfm{a}_{\rm A}\equiv
f_{\rm A} \bfm{B}$, where $f_{\rm O}\equiv \sqrt{r_{\rm O}}/B$,
$f_{\rm H}\equiv r_{\rm H}/B$, $f_{\rm A}\equiv \sqrt{r_{\rm
A}}/B$. $r_{\rm O}$, $r_{\rm H}$ and $r_{\rm A}$ are the Ohmic, Hall and 
ambipolar resistivities respectively and are given by
\begin{eqnarray}
r_{\rm O} & = & \frac{1}{\sigma_{\rm O}} , \\
r_{\rm H} & = & \frac{\sigma_{\rm H}}{\sigma_{\rm H}^2 +
	\sigma_{\rm A}^2} , \\
r_{\rm A} & = & \frac{\sigma_{\rm A}}{\sigma_{\rm H}^2 + \sigma_{\rm A}^2} ,
\end{eqnarray}
\noindent with the conductivities given by
\begin{eqnarray}
\sigma_{\rm O} & = & \frac{1}{B}\sum_{i=2}^N \alpha_i \rho_i \beta_i , \\
\sigma_{\rm H} & = & \frac{1}{B}\sum_{i=2}^N
						\frac{\alpha_i \rho_i}{1+\beta_i^2} ,\\
\sigma_{\rm A} & = & \frac{1}{B}\sum_{i=2}^N \frac{\alpha_i \rho_i 
	\beta_i}{1+\beta_i^2} ,
\end{eqnarray}
\noindent where $\beta_i$ is the Hall parameter for species~$i$ and is
given by
\begin{equation}
\beta_i = \frac{\alpha_i B}{K_{1\,i}\rho_1} .
\label{eqn-hallpar}
\end{equation}

As noted by \citet{falle03} and \citet{osd06}, the main difficulty with 
standard numerical techniques for integrating equation (\ref{B_eqn}) lies with 
the Hall term.  As this term becomes dominant the stable time-step goes to
zero.  However, O'Sullivan \& Downes (\citeyear{osd06, osd07}) presented a 
novel, explicit numerical method for integrating this term such that the limit 
on the stable time-step is not overly restrictive.  We use this 
``Hall Diffusion Scheme'' in this work.  Of course, all explicitly differenced 
diffusion terms give rise to a stable time-step which is proportional to 
$\Delta x^2$, where $\Delta x$ is the resolution of the simulation.  To 
ameliorate this we use standard subcycling of the Hall terms and super 
time-stepping to accelerate the ambipolar diffusion terms 
\citep[see][]{ale96, osd06, osd07}.

Equations (\ref{mass}) -- (\ref{B_eqn}) are solved using a standard
shock-capturing, second order, finite volume, conservative scheme.
The numerical techniques employed in this work are slightly different to those 
used in Paper I in one respect: we have altered the calculation of the 
advective fluxes in equation (\ref{B_eqn}) to use the method suggested
by \citet{falle03}. In Paper~I these fluxes were derived from interface 
values of the neutral gas velocity and the magnetic field. We find that at 
high resolutions with variable resistivities the latter method is prone to 
introducing grid scale features in the solution while the former is not. This 
undesirable effect was not an issue for the investigations carried out in 
Paper~I since resistivities were fixed. The downside of the described 
variation between the numerical approaches is that it must be considered as 
a possible source of discrepancy in comparisons between the results of 
Paper~I and this work. However, in order to provide evidence of the small 
influence, we have also run a fixed resistivity simulation for this work (see 
section \ref{subsec:initial-conditions}).

Equation (\ref{divB}) is enforced using the method of \citet{dedner02}.  The 
effects of the diffusive terms in equation (\ref{B_eqn}) are then incorporated 
in an operator split fashion.

\subsection{Initial conditions}
\label{subsec:initial-conditions}
%
%
We examine the decay of MHD turbulence in conditions suitable for 
dense regions of molecular clouds.  The conditions we use are similar to
those used in Paper I.  We briefly review them here for completeness.

The computational domain is set up as a cube of side $L=0.2$\,pc.  Periodic 
boundary conditions are enforced on all faces of the simulation 
domain.  The sound speed is set to 0.55\kms,  the initial density is chosen 
to be uniform with a value of $10^6$\pcm~and the magnetic field is also 
initially uniform in the $(1,1,1)$ direction with a magnitude of 1\,mG.  For 
these conditions, suitable conductivities are $\sigma_{\rm O} = 1\times10^{10}$
\,s$^{-1}$, $\sigma_{\rm H} = 10^{-2}$\,s$^{-1}$ and 
$\sigma_{\rm A} = 10^{-1}$\,s$^{-1}$ \citep{wardle99}.  We choose a 3-fluid 
set-up for our multifluid simulation: 1 neutral species and 2 charged 
species.  The densities, charge-to-mass ratios and collisional 
coefficients of the charged species are chosen in order to achieve these 
conductivities.  We choose these particular physical conditions with a view 
to maximizing the influence of the Hall effect in our simulations 
\citep{wardle99}.  In this way we hope to determine whether the 
Hall effect is ever likely to be important in molecular cloud turbulence.

The initial velocity field is defined to be the sum of waves with 64 
wave-vectors each with random amplitude and phase - i.e. 
\begin{equation}
q_{\alpha} = \sum_{j=0}^{64} A_{\alpha, j}
\cos(\bfm{k}_j\cdot\bfm{x}+ \phi_{\alpha, j})
\end{equation}
where $\alpha$ defines the component ($x$, $y$ or $z$) of the
appropriate quantity, $A_{\alpha, j}$ and $\phi_{\alpha, j}$ are the random 
amplitudes and phases and $\bfm{x}$ is the position vector.  We restrict the 
velocity field to be solenoidal (i.e.\ non-compressional). By construction the 
mean velocity over the domain is zero.

Table \ref{table:sim-defs} presents a complete list of the various 
simulations carried out in this work. The nomenclature we employ in 
referencing the simulations is xx-c where xx denotes the type of physics 
(e.g.\ a standard molecular cloud run is ``mc'', ideal MHD is ``mhd'' etc) 
and c is the resolution used.  The initial root-mean-square (rms) of the 
field is chosen to be 5 with a corresponding Alfv\'enic Mach number of
approximately 1.9.  In addition to the 4 multifluid MHD simulations run at 
different resolutions, we also run 4 further simulations.  The first is an 
ideal MHD simulation (mhd-512) which we use for comparison purposes and the 
other two (ambi-512 and hall-512) only incorporate one of ambipolar 
diffusion or the Hall effect, respectively. The final case (fr-512) is a 
fixed resistivity simulation used to make contact with the simulations of 
Paper~I. We use these latter 4 simulations to investigate separately the 
influence of each non-ideal effect.

\begin{table*}
\begin{center}
\caption{Definition of the initial conditions used in the simulations in
this work. \label{table:sim-defs}}
\begin{tabular}{lccl}
\tableline \tableline
Simulation & Mach number\tablenotemark{a} & Resolution & Comment \\
\tableline
mc-64 & 5 & $64^3$ & - \\
mc-128 & 5 & $128^3$ & - \\
mc-256 & 5 & $256^3$ & - \\
mc-512 & 5 & $512^3$ & - \\
ambi-512 & 5 & $512^3$ & $r_{\rm H}=0$ \\
hall-512 & 5 & $512^3$ & $r_{\rm A}=0$ \\
mhd-512 & 5 & $512^3$ & Ideal MHD \\
fr-512 & 5 & $512^3$ & As mc-512 but for fixed resistivities \\
\tableline
\end{tabular}
\tablenotetext{a}{Initial rms Mach number of the flow}
\end{center}
\end{table*}

\section{Analysis}
%
%
\label{sec:analysis}
In this section we discuss the method of analysis of the output of the
simulations described in section \ref{subsec:initial-conditions}.  The
aim of this paper is to investigate the decay rate of supersonic turbulence 
in molecular clouds.  Hence, the main analysis carried out is of the kinetic, 
magnetic and total energy as functions of time.  These quantities are defined 
respectively as
\begin{mathletters}
\begin{eqnarray}
e_{\rm k} & = & \int_{\rm domain} \rho |\bfm{q}|^2\,dV \label{kin-def}\\
e_{\rm b} & = & \int_{\rm domain} \frac{|\bfm{B}|^2}{2}\,dV - \frac{<\bfm{B}>^2}
{2}V
\label{b-def}\\
e_{\rm tot} & = & e_b + e_k \label{etot-def}
\end{eqnarray}
\end{mathletters}
where $V$ is the volume of the computational domain.  

We also calculate the mass-weighted average Mach number, defined by
\begin{equation}
M = \frac{1}{a}\left\{\sigma_x^2 + \sigma_y^2 + \sigma_z^2\right\}^{1/2}
\end{equation}
where $a$ is the sound speed and the velocity dispersions,
$\sigma_\alpha$, are defined by
\begin{equation}
\sigma_\alpha = \left\{\frac{\left<\rho
	q_\alpha^2\right>}{\left<\rho\right>}\right\}^{1/2}
\end{equation}
where $\alpha$ is either $x$, $y$ or $z$ and the angle brackets denote
averaging over the computational domain (see \citealt{lem09}).

In section \ref{sec:power-spectra} we present the power spectra for the 
velocity, density and magnetic field for each of the $512^3$ simulations.  
These spectra are calculated by taking the power spectra in the $x$, $y$ and 
$z$ directions separately and summing the power over the interval $k \leq 
|\bfm{k}| < k + \Delta k$ (where we take $\Delta k=1$).  This gives us some 
insight into the scale of structures being formed by the turbulence for the 
various initial conditions and range of physics examined.

Finally, in section \ref{sec:vel-disp} we calculate the velocity
dispersion as a function of length scale, $l$.  For these purposes we 
define the velocity dispersion to be
\begin{eqnarray}
\sigma(l) = \left\{<\sigma_x^2(l)>_{\rm domain} + <\sigma_y^2(l)>_{\rm
	domain} \nonumber \right. \\
\left.	
	+ <\sigma_z^2(l)>_{\rm domain}\right\}^{\frac{1}{2}}
\end{eqnarray}
\noindent where
\begin{equation}
\sigma_\alpha(l) = \left\{\left<q_\alpha^2\right>_l-\left< q_\alpha
\right>_l^2\right\}^{\frac{1}{2}}
\end{equation}
where $<\cdot>_l$ indicates an average taken over a cube of side $l$ in
the simulation domain and $<\cdot>_{\rm domain}$ indicates averaging of the
quantity over all such non-overlapping cubes within the domain.

\section{Results}
\label{sec:results}

Each of the simulations detailed in table \ref{table:sim-defs} was run for one 
sound crossing time, $t_{\rm c} = 3.56\times10^{4}$\,yrs, of the simulation 
domain.  All analysis was carried out for $t \geq 0.2$\,$t_{\rm c}$ (i.e.\ 
one flow crossing time) at which point we expect significant turbulent mixing 
to have taken place and the system's memory of the initial state to be largely 
forgotten. 

\begin{figure}
\epsscale{1.10}
\plotone{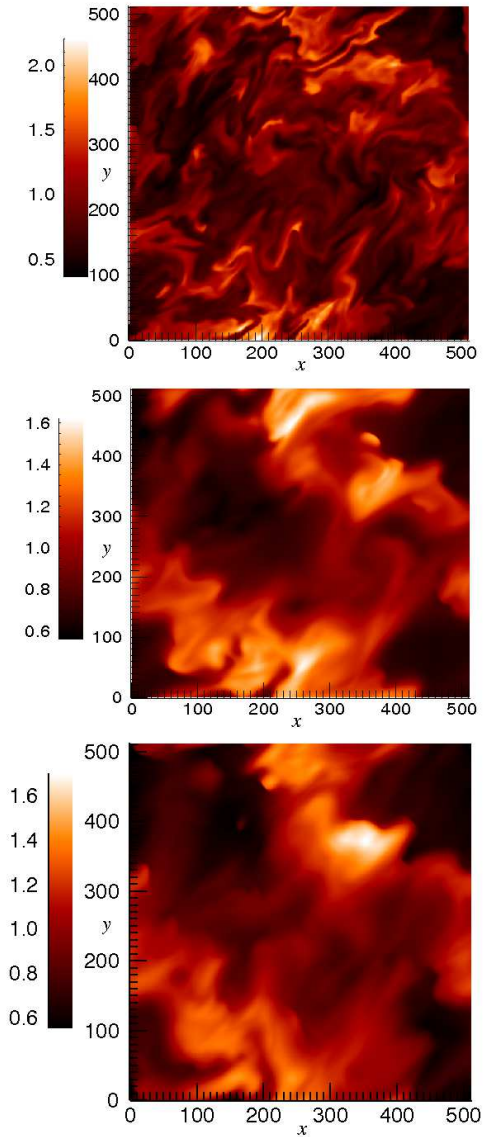}
\caption{Log plot of the neutral mass density at $t = t_{\rm c}$ in a slice at 
$z = 0.5$ for simulations mhd-512 (upper panel), mc-512 (middle panel) 
and fr-512 (lower panel). \label{fig:mhd-p2pc-fix-comp}}
\end{figure}

As an illustration of the differences between an ideal MHD turbulence
simulation and a multifluid MHD simulation, figure
\ref{fig:mhd-p2pc-fix-comp} contains plots of the density distribution
at $t=t_{\rm c}$ in a slice through the computational domain for simulations 
mhd-512 and mc-512.  It is clear that there is much less fine structure in 
mc-512.  Also shown is the same slice for simulation fr-512 (i.e.\ fixed 
resistivities).  While the similarities in terms of the levels of structure 
are relatively small between mc-512 and fr-512, there are clear differences 
between the distributions indicating that calculating the resistivities
self consistently has some impact on the dynamics of the system.

\begin{figure}
\epsscale{1.10}
\plotone{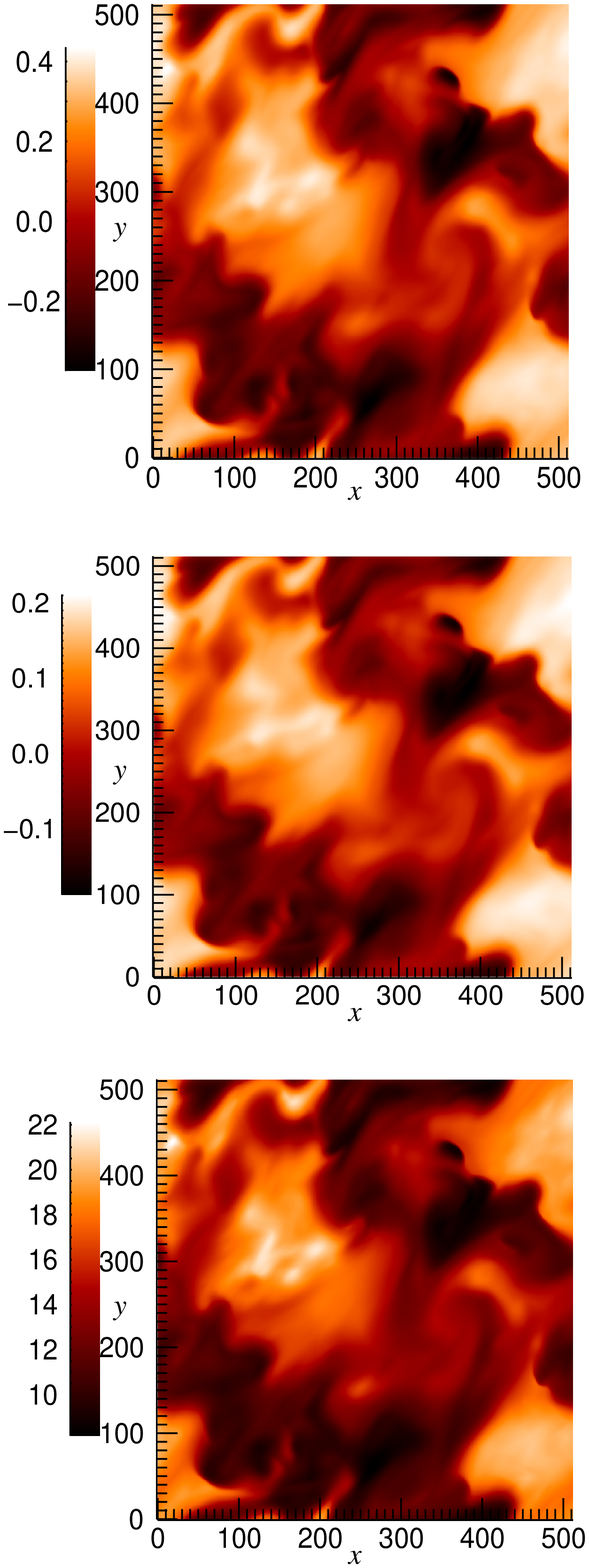}
\caption{Plot of the normalized ambipolar resistivity (top panel), normalized 
Hall resistivity (middle panel) and 
$\eta\equiv \frac{r_{\rm A}}{|r_{\rm H}|}$.  Note that the scale is 
logarithmic in the top and middle panels while it is linear for the bottom 
panel. \label{fig:p2pc-res}}
\end{figure}

\begin{figure}
\epsscale{1.10}
\plotone{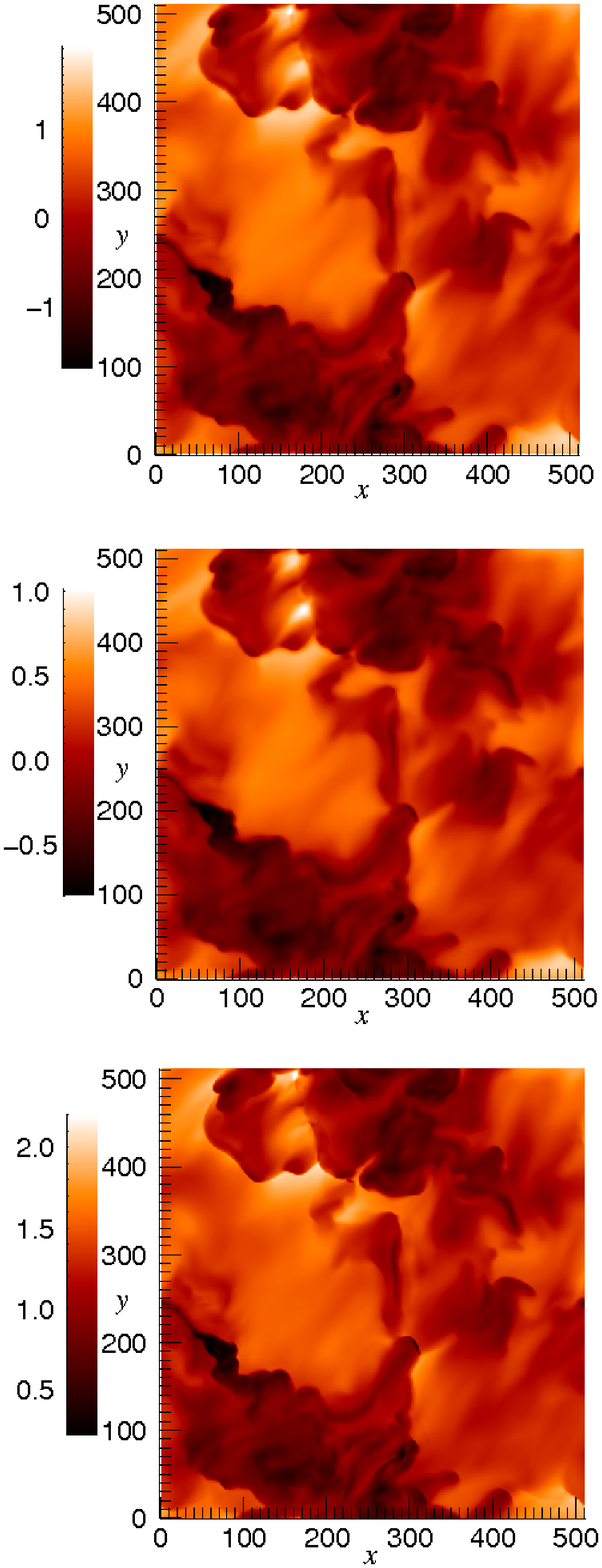}
\caption{As figure \ref{fig:p2pc-res} but the snapshots are taken at $t
= 0.2$\,$t_{\rm c}$.  Note that the scale here is logarithmic in each
case including the bottom panel.\label{fig:p2pc-res-early}}
\end{figure}

In figure \ref{fig:p2pc-res} we present plots of the ambipolar and Hall 
resistivities for simulation mc-512 for the same times and slices as figure 
\ref{fig:mhd-p2pc-fix-comp}.  It is clear that the resistivities vary
considerably throughout the computational domain with the features
strongly correlated with the features in the density distribution.  We
also show $\eta \equiv \frac{r_{\rm A}}{|r_{\rm H}|}$ to give an
indication of the relative importance of each of the resistivities.
This, as we shall see, is an important parameter.  Finally, figure
\ref{fig:p2pc-res-early} is the same as \ref{fig:p2pc-res} except that
the data is taken at time $t = 0.2 t_{\rm c}$ - i.e.\ after one flow
crossing time.  Here we can see that the variation of the resistivities
in space is dramatic with, for example, the ambipolar resistivity
varying by almost 4 orders of magnitude with $\eta$ varying by around 2 orders 
of magnitude.

\subsection{Resolution study}
\label{subsec:resolution-study}

\begin{figure}
\epsscale{1.00}
\plotone{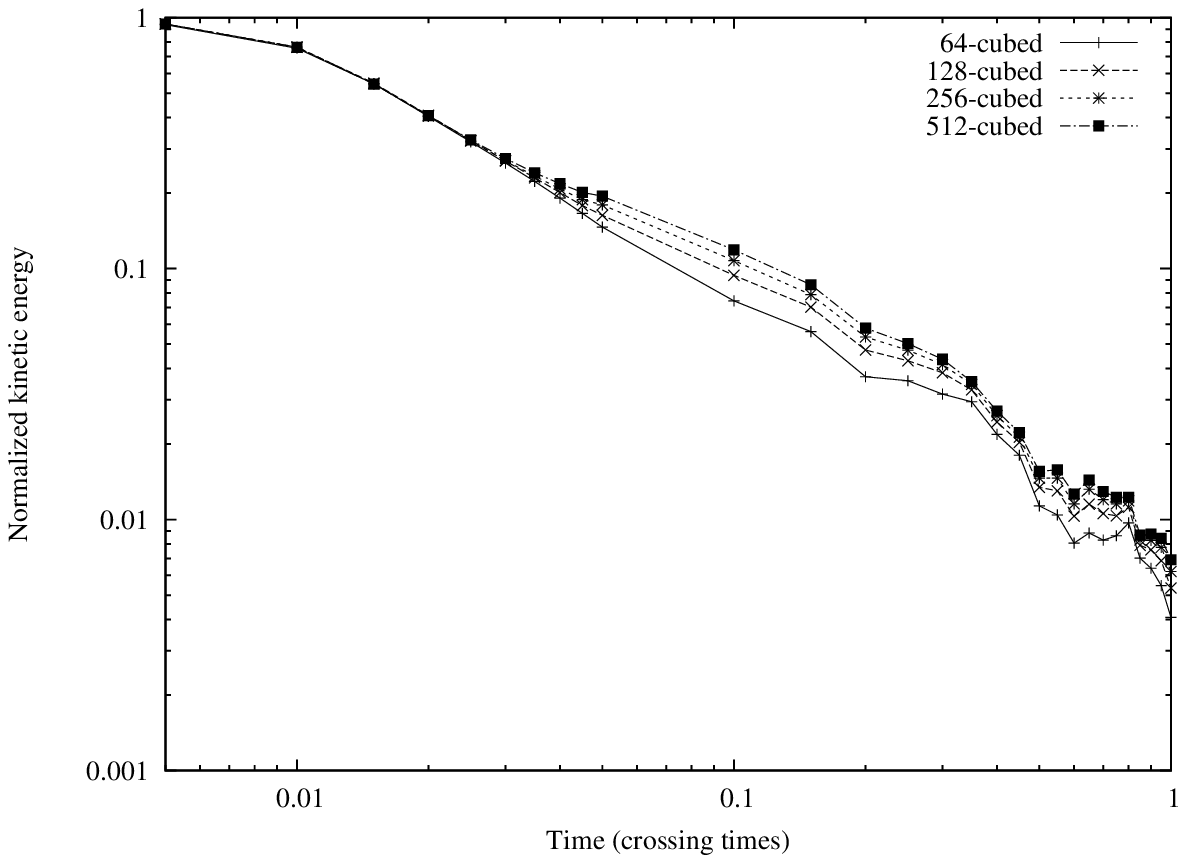}
\caption{Log scale plot of the kinetic energy (normalized to its initial
value) for each of the simulations in the resolution 
study.\label{fig:res-energy}}
\end{figure}

Four simulations identical in every way except for the resolution were run.  
Specifically, the resolutions used were $64^3$, $128^3$, $256^3$, and 
$512^3$.  We now focus our attention on how the energy decay behaves with 
resolution.  Figure \ref{fig:res-energy} contains plots of the kinetic energy 
as a function of time for each of the simulations in the resolution study.  It 
is clear that the lower the resolution, the faster the decay - this is what
one would expect since lower resolution results in a higher numerical 
viscosity and hence one expects faster dissipation of energy.  

Simulations mc-256 and mc-512 are, however, quite similar in terms of the 
energy decay with a maximum relative difference of around 10\% between the 
kinetic energies in the simulations at any one time - an almost identical 
result to that obtained from the resolution study in Paper I.  This is 
notable since in Paper I the resistivities were kept constant in space and 
time whereas here the resistivities locally increase significantly during the 
course of the simulations.  A reasonable inference is that the influence of 
local variations in resistivities averages out in some sense on the global 
scale.

As we shall see later, however, the effect of spatially varying resistivities
is noticeable in properties such as the power spectrum of the density
and magnetic field.

The various energy decay rates can be modeled approximately as power-laws 
in time, i.e.\ $t^{-\beta}$.  Fitting the kinetic energy, $\beta_{\rm K}$, 
magnetic energy, $\beta_{\rm B}$, and total energy, $\beta_{\rm Tot}$,
as functions of time in this way we obtain the values given in table 
\ref{table:decay-exp}.  This data confirms quantitatively what can be observed 
in Figure \ref{fig:ek-decay} and extends it to the decay of the energy
in magnetic perturbations: increasing resolution decreases the rate of energy 
decay, but the difference between the $256^3$ and $512^3$ simulations is 
relatively minor.  We note in passing that the decay in the energy in
magnetic perturbations is considerably more sensitive to resolution than
kinetic energy: $\beta_{\rm B}$ varies between 1.43 and 1.28 while
$\beta_{\rm K}$ only varies between 1.38 and 1.34.  We know that the
resistivities are a critical factor in determining the decay of the
magnetic energy since they facilitate loss of magnetic energy through
reconnection.  We attribute the extra sensitivity to resolution of $\beta_{\rm
B}$ to the necessity to properly resolve the diffusive effects in the
induction equation, {\em including} the variation of the resistivities
themselves.  The variation of the resistivities throughout the
computational domain is significant (see figure \ref{fig:p2pc-res}) and
it is interesting to note that $\beta_{\rm B}$ in simulations with fixed
resistivities is not so sensitive to resolution (see Paper I).

\begin{table}
\begin{center}

\caption{The values of the exponent for the kinetic, magnetic and total 
energy decay for the simulations presented in this work.  These
exponents are calculated by fitting the data over the time interval
$[0.2 t_{\rm c}, t_{\rm c}]$.
	\label{table:decay-exp}}
\begin{tabular}{lccc}
\tableline \tableline
Simulation & $\beta_{\rm K}$ & $\beta_{\rm B}$ & $\beta_{\rm Tot}$ \\
\tableline
mc-64 & 1.38 & 1.43 & 1.39 \\
mc-128 & 1.37 & 1.35 & 1.36 \\
mc-256 & 1.35 & 1.30 & 1.33 \\
mc-512 & 1.34 & 1.28 & 1.32 \\
hall-512 & 1.12 & 1.05 & 1.09 \\
ambi-512 & 1.35 & 1.30 & 1.33 \\
mhd-512 & 1.12 & 1.06 & 1.10 \\
fr-512 & 1.30 & 1.28 & 1.30 \\
\tableline
\end{tabular}
\end{center}
\end{table}

\subsection{Energy decay}
\label{sec:energy-decay}

We now discuss the behavior of the kinetic and magnetic energy in our
multifluid simulations and compare with those in Paper I.

\subsubsection{Kinetic energy decay}
\label{sec:kin-decay}

\begin{figure}
\epsscale{1.00}
\plotone{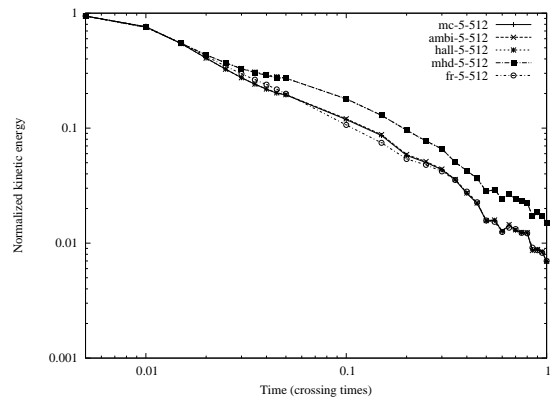}
\caption{Log scale plot of the kinetic energy (normalized to its initial
value) for each of the $512^3$ simulations.\label{fig:ek-decay}}
\end{figure}

Figure \ref{fig:ek-decay} contains plots of the decay of kinetic energy
with time for the $512^3$ resolution simulations outlined in table 
\ref{table:sim-defs}.  Note that the kinetic energy decay in all of the 
simulations is very similar until around $t \approx 0.02\, t_{\rm c}$.  This 
is because at such early times compressions are only just starting to form and 
so the non-ideal terms in the induction equation have had almost no effect on 
the dynamics.  The subsequent energy decay of simulations mhd-512 and hall-512
are almost identical to each other.  The energy decay of simulations mc-512 
and ambi-512 are virtually identical over the full plotted range while the 
data plotted for fr-512 coincide with the former simulations for times
in the range $0.03\,t_{\rm c} \leq t \leq 0.2\,t_{\rm c}$.  We have fitted 
the kinetic energy decay as a power-law in the range 
$0.2 t_{\rm c} \leq t \leq t_{\rm c}$ (i.e.\ after approximately one
initial flow crossing time) and the exponents are given in the first column of 
table \ref{table:decay-exp}.  

It is clear that the presence of ambipolar diffusion has a significant impact 
on the behavior of the kinetic energy in the turbulent system.  This is a 
result of the exchange of energy between kinetic and magnetic energies 
as will be discussed in section \ref{sec:b-decay}.

\begin{figure}
\epsscale{1.00}
\plotone{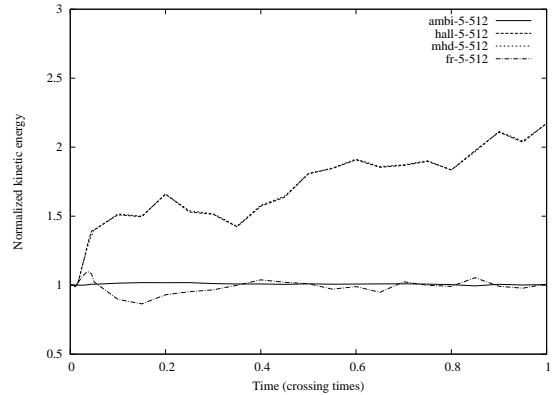}
\caption{Plot of the ratio of the kinetic energy in our
$512^3$ simulations to that in mc-5-512\label{fig:ek-p2pc-ref}}
\end{figure}

From figure \ref{fig:ek-decay} and table \ref{table:decay-exp} it is
evident that the Hall effect has almost no impact on the kinetic energy decay 
in turbulence in molecular clouds.  In order to emphasize any possible
impact of the Hall effect we have plotted the time evolution of the ratio of 
the kinetic energy in each of our simulations to that in mc-5-512 in 
figure \ref{fig:ek-p2pc-ref} on a linear scale.  It is clear even in
this figure that the Hall effect has little impact on the evolution of
the turbulence.  This result is also reproduced if we examine the
evolution of the magnetic energy (see section \ref{sec:b-decay}).  This 
supports our conclusion from Paper I in which the simulations were run using 
fixed resistivities (see also fr-512 in this work).

Also shown in figure \ref{fig:ek-decay} is the energy decay for simulation 
fr-512.  Given the wide variation of the resistivities in both space
and time (see figures \ref{fig:p2pc-res} and \ref{fig:p2pc-res-early}) it is
somewhat surprising that the energy decay is so similar to that of
mc-512.  In fact, the volume average of the ambipolar resistivity at 
$t = 0.5$\,$t_{\rm c}$ in mc-512 is approximately 60\% {\em higher} than that 
in the fr-512 simulation.  It would appear that, while ambipolar diffusion 
enhances energy loss, the expected spatial and temporal variation of it does 
not have that much influence.

\begin{figure}
\epsscale{1.10}
\plotone{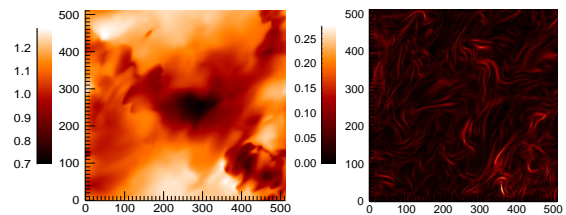}
\caption{Slices showing the $z=0.5$ plane at $t=0.5$\,$t_{\rm c}$ for both
$|\mathbf{B}|$ (left panel) and $\delta B$ (right panel).  \label{fig:magb-db}}
\end{figure}

One possible reason for the small divergence between mc-512 and fr-512 
would be if the locations in which the resistivity is high are regions in 
which the magnetic field, $\mathbf{B}$, is varying weakly.  To explore this we 
define a scalar, $\delta B$, by
\begin{equation}
\label{db-def}
\delta B \equiv |\grad(B_x)| + |\grad(B_y)| + |\grad(B_z)|
\end{equation}
\noindent where $\grad(B_x)$, for example, is a normalized gradient defined by
\begin{equation}
\grad(B_x) \equiv \frac{1}{B_0} \left( \begin{array}{c} \delta_x B_x \\
		\delta_y B_x \\
		\delta_z B_x \end{array} \right)
\end{equation}
where $\delta_\cdot$ means centered differencing in the indicated
direction {\em without} normalizing by the zone spacing and $B_0$
is the magnitude of the magnetic field throughout the domain at $t=0$.
$\delta B$ is then a dimensionless measure of the variation of 
$\mathbf{B}$ at any point in space or time: if $\delta B$ is large it means 
there is a high gradient in one or more of the components of $\mathbf{B}$ and 
therefore resistivity will have an important influence here. Figure 
\ref{fig:magb-db} contains snapshots of $\delta B$ and $|\mathbf{B}|$ at 
$t=0.5$\,$t_{\rm c}$. There is rich structure in $\delta B$ which is not 
apparent in $|\mathbf{B}|$.

\begin{figure}

\epsscale{1.10}
\plotone{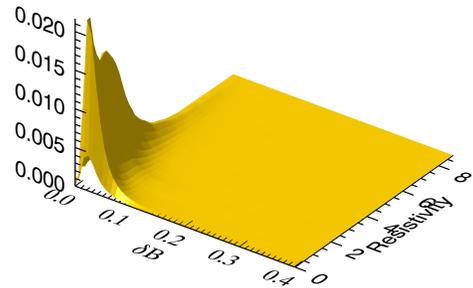}
\caption{Probability density function for the normalized variation in 
$\mathbf{B}$ and the resistivity normalized to its initial value
(i.e.\ the same as that used in simulation fr-512).  It is clear that high 
resistivities only occur where there is low variation in 
$\mathbf{B}$.  \label{fig:db-ambi}}
\end{figure}

Figure \ref{fig:db-ambi} contains a histogram plot of the two
dimensional probability density function for $\delta B$ and $r_{\rm A}$
(the ambipolar resistivity).  What is striking about this plot is that
there is a notable lack of high $\delta B$ with corresponding high
resistivity.  Of course, if we have any system in which there are
regions of high and low resistivity we expect that, over time, the regions
with high diffusion will have lower variation in $\mathbf{B}$ so this, in 
itself, doesn't tell us much.  However, it does prompt us to look a little 
more closely at the behavior of $r_{\rm A}$.

Comparison of the middle panel of figure \ref{fig:mhd-p2pc-fix-comp} with the 
top panel of figure \ref{fig:p2pc-res} shows that the ambipolar resistivity is 
higher in regions of low density, as would be expected.  Now, in
supersonic/super-Alfv\'enic turbulence kinetic energy is dissipated most
strongly at strong shocks.  However, shocks propagating into regions of
very low density will not dissipate kinetic energy effectively.  Since it is
precisely these regions in which our resistivities are high we must
conclude that, in fact, the regions of enhanced resistivity do not
contribute significantly to energy dissipation and hence we would not
expect the introduction of spatially varying resistivities to increase
the rate of energy decay in our simulations.  Further, since the volume
average of, for example, the ambipolar resistivity is actually higher
than that used in fr-512 we would not expect the spatial variation of
this resistivity to reduce the rate of energy decay either.

\subsubsection{Magnetic energy decay}
\label{sec:b-decay}

We now move on to discuss the decay in magnetic energy.  Initially, as
outlined in Paper I, the magnetic energy increases as the flow
compresses and stretches the magnetic field throughout the computational
domain.  Once this initial increase in the energy has occurred it
is gradually lost through two main avenues: magnetic reconnection and transfer 
of magnetic energy to kinetic energy which can then be dissipated in shocks 
and other viscous processes.

\begin{figure}
\epsscale{1.00}
\plotone{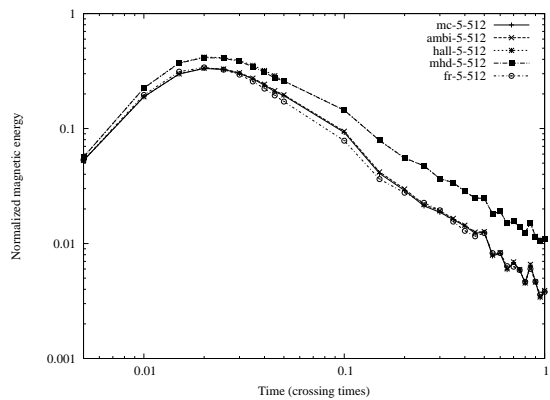}
\caption{As figure \ref{fig:ek-decay} but for magnetic energy.
\label{fig:eb-decay}}
\end{figure}

Figure \ref{fig:eb-decay} contains plots of the decay of magnetic energy
with time.  It is clear that, in common with the case of kinetic energy, the 
hall-512 and mhd-512 simulations are almost identical while the ambi-512 and 
mc-512 simulations are also well matched.  This supports our inference from 
section \ref{sec:kin-decay} that the Hall effect has almost no impact on 
energy decay in molecular clouds on the global scale.

Ambipolar diffusion does, however, have a significant impact on the
behavior of both the kinetic and magnetic energies.  This was also noted
in Paper I and we explain it in the same way, recapitulated briefly here for
completeness.  Consider a region of the flow undergoing compression.  During
this compression kinetic energy will be converted into both magnetic and
internal energy through increasing the magnetic pressure and the thermal
pressure.  Given that this is a turbulent flow we expect that, after
some time, this region will begin to expand again.  However, during the
compression ambipolar diffusion will have diffused away some of the
magnetic energy thereby leaving less to be converted back to kinetic
energy.  In this way the presence of ambipolar diffusion creates a new path
through which energy can be lost from the system and reduces the level
of all forms of energy in the system.

\subsection{Power spectra}
\label{sec:power-spectra}

We now move on to a study of the power spectra obtained from the
multifluid MHD turbulence simulations.  These spectra are important from
the point of view of understanding the types of structures formed by the
turbulence and are a more discerning tool for exploring any structural 
differences caused by multifluid effects.  Table \ref{table:power-slopes} 
contains the exponents of the power spectra assuming a power-law relationship 
between power and wave number.  All the analysis presented in this section is 
performed on data taken at $t=t_{\rm c}$.

\begin{table*}
\begin{center}
\caption{The values of the exponent for the power spectra of density,
velocity and magnetic field measured at $t = t_{\rm c}$.  All fits are
over the range $5 \leq k \leq 20$ unless otherwise noted.
\label{table:power-slopes}}
\begin{tabular}{lccc}
\tableline \tableline
Simulation & Density & Velocity & Magnetic field \\ 
\tableline
mc-512 & 1.82\tablenotemark{a}, 4.33\tablenotemark{b} & 1.34
& 2.14\tablenotemark{a}, 5.38\tablenotemark{b} \\
ambi-512 & 1.79\tablenotemark{a}, 4.31\tablenotemark{b} &
1.34 & 2.15\tablenotemark{a}, 5.43\tablenotemark{b}\\
hall-512 & 1.25 & 1.01 & 1.51 \\
mhd-512 & 1.27 & 1.00 & 1.55 \\ 
fr-512 & 2.11\tablenotemark{a}, 4.21\tablenotemark{b} & 1.28
& 2.20\tablenotemark{a}, 5.53\tablenotemark{b} \\
\tableline
\end{tabular}
\tablenotetext{a}{Fitted over $4 \leq k \leq 10$}
\tablenotetext{b}{Fitted over $10 \leq k \leq 100$}
\end{center}
\end{table*}

\subsubsection{Density power spectra}
\label{sec:rho-power}

\begin{figure}
\epsscale{1.00}
\plotone{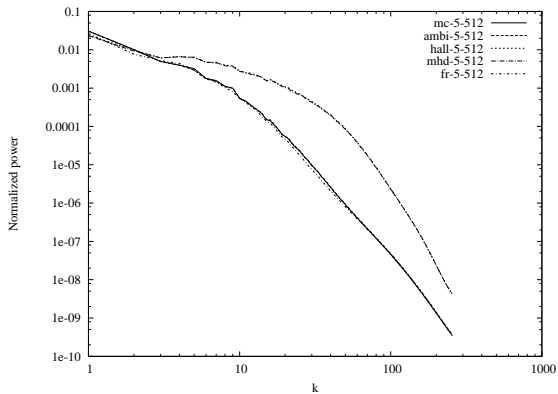}
\caption{Plot of the normalized, spherically integrated power spectrum
of the neutral mass density for each of the $512^3$ simulations at 
$t=t_{\rm c}$. \label{fig:rho-power}}
\end{figure}

We turn first to the scales of the structures formed in the density
distributions for our various simulations.  Figure \ref{fig:rho-power}
contains plots of the power spectra of the neutral density (or density,
in the case of simulations mhd-512 and fr-512) for all the $512^3$
resolution simulations.  

The power spectra for the simulations including the effects of ambipolar 
diffusion are approximately broken power laws made up of 3 distinct power laws: 
$1\lesim k \lesim 10$, $10 \lesim k \lesim 100$ and $100 \lesim k$.  Below 
the low $k$ break, the spectrum is dependent on the scale of the computational 
domain. At high $k$ approaching the grid scale, numerical viscosity will begin 
to dominate.  In common with the results presented so far we see that there is 
little difference between simulations mc-512 and ambi-512 - in fact it
is difficult to distinguish between the two spectra without careful
examination of figure \ref{fig:rho-power}.  The fr-512 power spectrum is
also similar to mc-512 and ambi-512 although it has slightly less
power at intermediate values of $k$ with the difference here being at most 
10\%.

There is almost no detectable difference between simulations hall-512 and 
mhd-512.  Evidently, the Hall effect has a much weaker influence on the 
density structure in molecular clouds than ambipolar diffusion.  Ambipolar 
diffusion, on the other hand, has a very significant impact with pronounced 
damping of density structures at scales less than one tenth of the domain size
(corresponding to a physical scale of approximately 0.02\,pc).  This damping 
is evident in figure \ref{fig:mhd-p2pc-fix-comp} where the density structures 
in simulations with ambipolar diffusion are more smeared than in mhd-512.

From a quantitative perspective, the data in table \ref{table:power-slopes} 
shows that the inclusion of ambipolar diffusion significantly steepens the 
power spectra, increasing the exponent by more than 0.5 over the cases which 
do not include the effect.  The data again suggests that the Hall effect has 
minimal impact.  

The results for fr-512 show a softer spectrum at large length 
scales and a harder spectrum at short scales than mc-512 and ambi-512,
indicating that self-consistent calculation of the resistivities reduces
the level of fine structure.  This result must, however, be confirmed by
higher resolution simulations before it can be regarded as reliable.

\begin{figure}
\epsscale{1.00}
\plotone{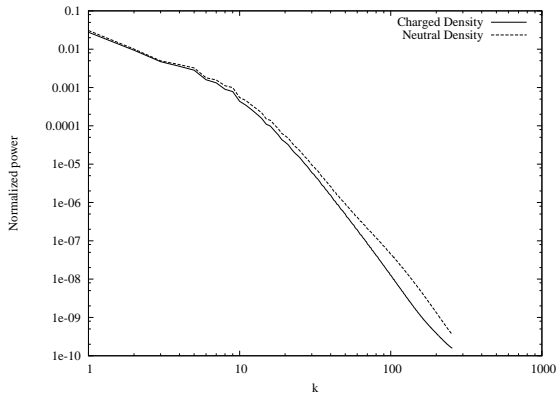}
\caption{Plot of the normalized, spherically integrated power spectrum
of the neutral and negatively charged mass densities for simulation
mc-512. \label{fig:rho-neutral-s1-comp}}
\end{figure}

Figure \ref{fig:rho-neutral-s1-comp} contains plots of the power spectra
of the neutral density and the density of the negatively charged species at
$t=t_{\rm c}$ for comparison.  It can be seen that the neutral mass
density has more power for $k \geq 10$ although the qualitative shape of
the power spectra are the same in each case.  This is in qualitative
agreement with the results presented in \cite{li08} for driven
turbulence simulations.  For comparison with the results in Table 
\ref{table:power-slopes}, the exponent for the charged species mass density in 
the range $4 \leq k \leq 10$ is -2.21 while in the range $10 \leq k \leq 100$ 
it is -4.84.  The power spectrum for the neutral mass density is harder than 
that for the magnetic field and, in the range $10 \leq k \leq 100$, the 
spectrum for the charged species mass density lies somewhere between 
the two.  This is not surprising as it is a function of the neutral
density and velocity (through drag) and the magnetic field (through the 
Lorentz force).

\subsubsection{Velocity power spectra}
\label{sec:vel-power}

\begin{figure}
\epsscale{1.00}
\plotone{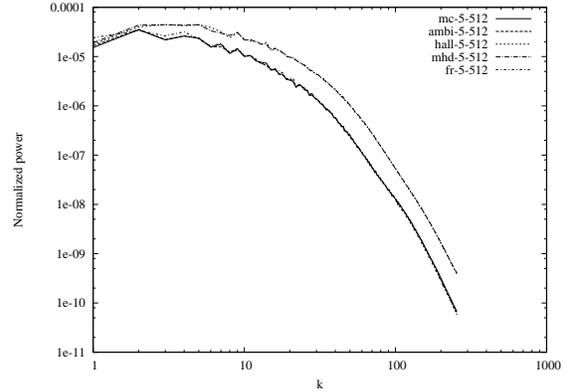}
\caption{Plot of the normalized, spherically integrated power spectrum
of the neutral velocity for each of the $512^3$ simulations at 
$t=t_{\rm c}$. \label{fig:v-power}}
\end{figure}

Figure \ref{fig:v-power} contains plots of the velocity power spectrum
for the neutral velocity.  In common with section \ref{sec:rho-power} those 
simulations incorporating ambipolar diffusion have significantly less power at 
almost all scales than those without.  This is to be expected given the 
increased rate of loss of turbulent energy in the presence of ambipolar 
diffusion (see section \ref{sec:energy-decay}).  

As found in Paper I, and again here, there are clear differences of a 
qualitative nature between the density power spectra and the velocity power 
spectra for the multifluid simulations.  Those simulations incorporating
ambipolar diffusion (mc-512 and ambi-512) exhibit a strong power-law
in the range $10 \leq k \leq 100$ in the density power spectrum.  This
is not true of the velocity power spectra.  For
the latter spectra there is a break at roughly $k \sim 20$ and again at
$k \sim 100$.  The latter break we interpret as being at the scale where 
numerical diffusive effects begin to dominate the non-ideal effects in the 
induction equation. The lower break can reasonably be interpreted as the
scale at which the non-ideal effects become important.  The marked
qualitative differences between the density and velocity power spectra 
indicate a considerable decoupling between the two variables.  It is worth
recalling here that the power spectra are calculated at $t=t_{\rm c}$
so it is reasonable to expect that the turbulence is well developed at this 
stage.  

In common with the density power spectra, the presence of ambipolar diffusion 
produces steeper velocity power spectra (see Table \ref{table:power-slopes}) 
in qualitative agreement with the results of \cite{li08}. 

The results at high $k$ are, of course, dominated by numerical diffusive 
effects.  However, it is interesting to note that simulation fr-512
actually has very slightly {\em less} power at short scales than mc-512 
despite the volume average of $r_{\rm A}$ being roughly twice as
high as the resistivity used in fr-512.  This adds weight to the
inference from section \ref{sec:kin-decay} that regions of high
resistivity tend not to be coincident with regions of high gradients in 
the magnetic field and therefore do not have the level of influence one
would naively expect.

\begin{figure}
\epsscale{1.00}
\plotone{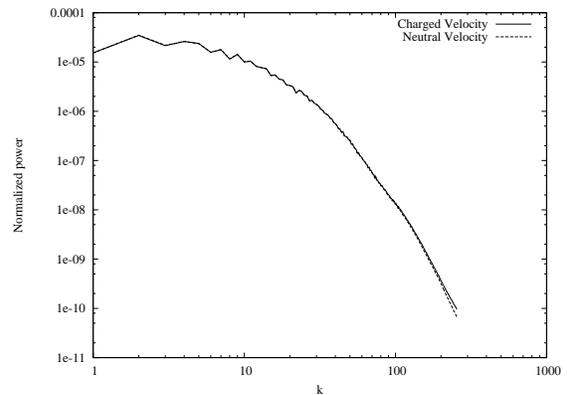}
\caption{Plot of the normalized, spherically integrated power spectrum
of the neutral velocity and the velocity of the negatively charged
species for simulation mc-512 at $t=t_{\rm c}$.
\label{fig:v-neutral-s1-comp}}
\end{figure}

Figure \ref{fig:v-neutral-s1-comp} contains plots of the power spectra
for the velocity of the neutral and negatively charged species.  It can
be seen that, except at very high $k$, they are virtually identical.
Given that the charged velocity is defined by balance between drag with
the neutrals and the Lorentz force this is, perhaps, unsurprising.

\subsubsection{Magnetic field power spectra}
\label{sec:b-power}

\begin{figure}
\epsscale{1.00}
\plotone{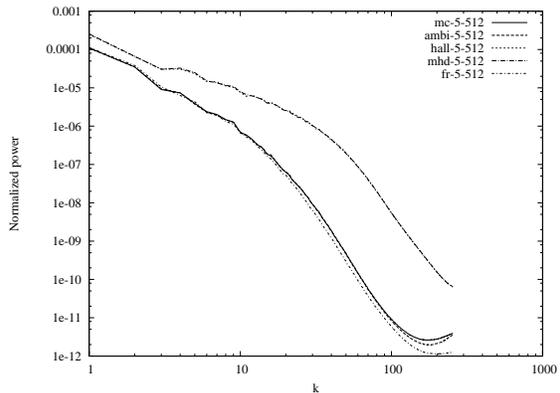}
\caption{As figure \ref{fig:v-power} but for the magnetic field.
	\label{fig:b-power}}
\end{figure}

Figure \ref{fig:b-power} contains plots of the spherically integrated
power spectra for the magnetic field.  Once again, ambipolar diffusion
has a much bigger impact on these spectra than the Hall effect.  The
magnetic field power spectra are considerably steeper at all $k$ in its
presence.  The absolute power at any scale is also considerably lowered
by ambipolar diffusion, as would be expected from the discussion in
section \ref{sec:b-decay}.  There is a qualitative similarity between
the density power spectra (figure \ref{fig:rho-power}) and the magnetic field
power spectra which is absent when comparing the latter with the
velocity power spectra (see figure \ref{fig:v-power}).

Interestingly, we see the phenomenon that fr-512 has very slightly
less power at $k \geq 20$ than mc-512 indicating that fixing the
resistivities at $t=0$ actually results in slightly more dissipation than
allowing it to vary in time and space.

We find that the Hall effect has a slightly more noticeable effect on
the magnetic field power spectra than on the spectra of velocity or
density: it gives rise to a little more structure on short scales which is 
absent in its absence.  This would be expected as the Hall effect is a 
dispersive effect acting directly on the magnetic field.  The results for the 
density and velocity power spectra, however, demonstrate that this influence 
over the magnetic power spectra does not translate into an influence over 
the other variables.  

\subsection{Velocity dispersion}
\label{sec:vel-disp}

\begin{figure}
\epsscale{1.00}
\plotone{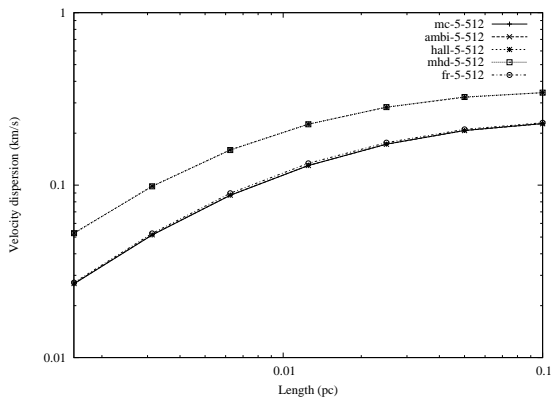}
\caption{Log-scale plot of the velocity dispersion as a function of
length scale for each of the $512^3$ simulations.  There is no power-law 
apparent in this data. \label{fig:vel-disp}}
\end{figure}

It has been widely reported that the observed velocity dispersion in
molecular clouds behaves as a power-law in the size of the field of
view.  Figure \ref{fig:vel-disp} contains plots of the velocity
dispersion for each of the $512^3$ resolution simulations presented in this 
work.  As noted in Paper I, and \cite{lem09}, no power-law is observed.  This
may be due to the fact that in (multifluid) MHD turbulence there
are many relevant signal speeds due to the variation in the orientation
and intensity of the magnetic field, in contrast to the situation with
hydrodynamic turbulence.  Hence there is no reason to expect to see a
power-law.  Finally, it should be noted that, at this point in the
simulation, the turbulence has decayed such that it is largely sub-sonic
or transonic.  This may also explain the lack of a power-law.  Indeed,
it should also be noted that our results here are not necessarily in
contradiction with observations since the velocity dispersion-size
correlation can only be accurately measured in the supersonic regime.

It is clear that ambipolar diffusion reduces the velocity dispersion at
all length scales - this is what would be expected given the results
from the velocity power spectra presented in section \ref{sec:vel-power}.
Once again the results here indicate that the Hall effect has little
impact.  Again, it is interesting to note that the spatial variation of
the resistivity in mc-512 appears to have almost no impact as the
velocity dispersion seen in fr-512 is almost identical to that in
mc-512.

\section{Conclusions}
\label{sec:conclusions}

We have presented results from a suite of $512^3$ resolution simulations of 
fully multifluid MHD decaying turbulence.  The effects incorporated include
the Hall effect and ambipolar diffusion.  We have performed a resolution
study to ensure that the energy decay rate, being the main result
presented here, is reliable.  We have confirmed the results of the
simplified calculations in Paper I that the Hall effect has little
impact on the nature and behavior of turbulence in molecular clouds
under the well motivated physical parameters assumed in this work.
Further, the presence of ambipolar diffusion increases the rate of
energy decay at length scales of 0.2\,pc and less.  The same conclusion
is drawn for the behavior of the energy in the magnetic field.

The power spectra for these simulations again suggest that the Hall
effect has little impact on the flows with the exception of the spectrum
of magnetic field variations.  We must keep in mind that the maximum 
resolution used here ($512^3$) is only enough to resolve about half a decade 
in $k$-space and it is therefore difficult to be confident of the details of 
the power spectra.  Notwithstanding this consideration it does appear clear 
that ambipolar diffusion steepens the power spectra of the neutral velocity, 
density and the magnetic field.  As noted in Paper I, it appears that at a 
resolution of $512^3$ and an assumed length scale of 0.2\,pc we have resolved 
the length at which ambipolar diffusion begins to influence the turbulent
cascade.  In Paper I only constant resistivities were implemented and
hence it was unclear whether this latter result would survive the inclusion 
of more realistic fully multifluid MHD in which the resistivities vary 
strongly in space and time.  The results presented here imply that it
does.

The power spectra of the neutral velocity and the magnetic field differ
qualitatively from that of the density with breaks occurring in the
former which are not seen in the latter.  This suggests a decoupling
between these fields.

The velocity dispersion as a function of length does not behave as a
power law.  This is not unexpected as the nature of MHD turbulence
implies a wide range of applicable signal speeds which can, when
combined, remove the power-law behavior which might be expected if only
one signal speed were relevant.

\acknowledgments{The authors wish to acknowledge the SFI/HEA Irish
Centre for High-End Computing (ICHEC) for the provision of computational 
facilities and support.  The work described in this paper was carried out 
using resources provided to ICHEC through the Irish National Capability 
Computing Initiative, a partnership between all the major third level research 
institutions and IBM coordinated by the Dublin Institute for Advanced Studies 
and supported by the HEA under PRTLI cycles 3 and 4 with funding from the 
ERDF and the NDP.  This work was partly funded by SFI through the
Research Frontiers Programme, grant number 07 PHYF586. Finally, the
authors would like to thank the anonymous referee for helpful
suggestions and comments on this work.}

\end{document}